\documentstyle[12pt]{article}
\begin{document}

\begin{center}
{\Large\bf Chiral Symmetry, Renormalization Group \\
\vspace*{3mm}
and Fixed Points for Lattice Fermions}
\footnote{Work supported by Conselho Nacional de Desenvolimento Cientifico
e Tecnologico}

\vspace*{1cm}

W. Bietenholz \\
Centro Brasileiro de Pesquisas Fisicas \\
rua Dr. Xavier Sigaud 150 \\
22290-180 Rio de Janeiro RJ \\
Brazil\\

\end{center}

\begin{abstract}
We discuss fixed point actions for various types of
free lattice fermions. The iterated block spin renormalization group
transformation yields lines of local but chiral symmetry breaking
fixed points. For staggered fermions at least the $U(1) \otimes U(1)$
symmetry can be preserved. This provides a basis for approximating perfect
actions for asymptotically free theories far from the critical surface.
For a class of lattice fermions that includes Wilson fermions we find
in addition one non local but chirally invariant fixed point. Its vicinity
is studied in the framework of the Gross Neveu model with weak
four Fermi interaction.
\end{abstract}

\vspace*{1cm}

To be published in Proceedings of the XIV Encontro Nacional de Fisica de
Particulas e Campos, Caxambu MG, Brazil (1993)

\pagebreak

In the following I report on a work done in collaboration with
U.-J. Wiese from HLRZ J\"{u}lich, Germany.\\

The naive action of free fermions on a hypercubic lattice of unit
spacing in $d$ dimensional Euclidean space reads:
\begin{equation} \label{snaiv}
S_{naive} = \frac{1}{2} \sum_{x} \sum_{\mu =1}^{d}[ \bar \psi_{x} \gamma_{\mu}
\psi_{x+\hat \mu} - \bar \psi_{x+\hat \mu} \gamma_{\mu} \psi_{x} ]
+ m \sum_{x} \bar \psi_{x} \psi_{x}
\end{equation}
The spinors are defined on the lattice sites $x$ ,
$\vert \hat \mu \vert =1$ and $\{ \gamma_{\mu},\gamma_{\nu} \} =
2 \delta_{\mu \nu}$.

The corresponding propagator in momentum space,
\begin{equation} \label{prop}
G(p) = [i \sum_{\mu} \gamma_{\mu} \sin p_{\mu} +m ]^{-1}
\end{equation}
displays the notorious ``fermion doubling'': for $m=0$ there are
poles in the Brillouin zone $B=]-\pi , \pi ]^{d}$
for $p_{\mu} =0$ {\em or} $p_{\mu} = \pi$, i.e. there appear
$2^{d}$ fermions instead of one. The additional fermions are an artifact
of the lattice; it is an outstanding problem how to get rid of them.
Hopes to realize this in a simple way were destroyed by the No Go
theorem of Nielsen and Ninomyia \cite{NN}.
The exact minimal assumptions for its
proof are complicated; we simplify them to:

1) Chiral invariance \ 2) Locality \ 3) Technicality ($G(p)$ has only
a finite number of poles in $B$)
\ 4) Unitarity \ 5) Translational
invariance of $S$ with respect to any lattice vector.\\
Then there are as many left- as righthanded fermions with the same internal
quantum numbers. In particular we can not put a single fermion on the
lattice.

There are many attempts to circumvent this theorem by violating
one of its assumptions and hoping that this violation disappears
in the conti-nuum limit. We give an (incomplete) list of them
\footnote{We ignore e.g. the Kaplan fermions, which are -- like
the staggered fermions -- not described by the ansatz (7), and the
Zaragoza fermions, one type of which is similar to the Stamatescu-Wu
fermions, see below. Moreover we don't discuss the attempts to solve the
problem by using random lattices.}
referring to the list of assumptions:

1) Wilson fermions \cite{Wilson}:
\begin{equation}
S_{Wilson} = S_{naive} + \frac{r}{2} \sum_{x,\mu} ( 2 \bar \psi_{x} \psi_{x}
- \bar \psi_{x} \psi_{x+\hat \mu} - \bar \psi_{x} \psi_{x-\hat \mu})
\end{equation}
If $p$ has $n$ components equal to $\pi $, then the mass increases as
$m \to m + 2nr$. Hence the unphysical poles disappear, but Wilson's term
breaks explicitly the chiral symmetry. The parameter $r$ is supposed to
disappear in the continuum, restoring the chiral symmetry.

2) SLAC fermions \cite{SLAC}: here a non locality is introduced by hand.
The sinus in (\ref{prop}) is linearized such that $G^{-1}(p)$
performs a finite gap at the boundary of $B$.
Also Rebbi fermions \cite{Rebbi} contain an artificial non locality. Here
$G^{-1}(p)$ has poles.

3) Smeared fermions \cite{Trivedi} include couplings to next to
nearest neighbors.
They are chirally symmetric, but violate technicality: $G^{-1}(p)$
vanishes on the whole boundary of $B$.

4) Stamatescu-Wu fermions \cite{SW}: take a one-sided difference for the
lattice
derivative instead of the symmetrized expression in (\ref{snaiv}).
The extra poles disappear, but unitarity is lost.

5) Staggered fermions \cite{Susskind}:
we decouple the flavors, most easily by the substitution \cite{KS}
\begin{displaymath}
\bar \psi_{x} \to \bar \psi_{x} \gamma_{1}^{x_{1}} \gamma_{2}^{x_{2}}
\dots \gamma_{d}^{x_{d}} \ ; \quad \psi_{x} \to \gamma_{d}^{x_{d}} \dots
\gamma_{1}^{x_{1}} \psi_{x} \ ,
\end{displaymath}
and then consider only one of them:
\begin{equation}
S_{stag} = \sum_{x} [ \frac{1}{2} \sum_{\mu} \sigma_{\mu}(x)
( \bar \chi_{x} \chi_{x+\hat \mu} - \bar \chi
_{x+\hat \mu} \chi_{x} ) + m \bar \chi_{x} \chi_{x} ]
\end{equation}
$\bar \chi , \ \chi $ are one component Grassmann fields and
the sign factor $
\sigma_{\mu}(x) \doteq (-1)^{x_{1}+ \dots + x_{\mu -1}} $
makes $S_{stag}$ explicitly $x$ dependent; it is only translational invariant
under an even number of lattice spacings. Here we
reduce the fermion multiplication to the factor
$2^{d/2}$.\\

For free fermions, the critical surface corresponds to the
chiral limit $m=0$. There we perform renormalization group transformations
(RGT) and hope to find a fixed point action (FPA). The FPA is free of
cutoff effects.

Kadanoff's block spin RGT acts like this:
\begin{equation}
e^{-S'[\bar \psi ',\psi ']} = \int D \bar \psi D \psi
K[\bar \psi ' , \psi ' , \bar \psi , \psi ] e^{-S[\bar \psi , \psi ]}
\end{equation}
$\bar \psi' , \psi' $ are new spinors defined on the centers of
hypercubic blocks, i.e. on a coarser lattice, and $S'$ is the transformed
action. This transformation must not change the partition function.
Thus we reduce systematically the number of degrees of freedom
by integrating out the short range fluctuations inside a block
(i.e. the high momentum modes).
We obtain a sequence $S,S',S'' $ etc. which might converge to an
FPA $S^{*}$. In the fermionic case, a sensible choice for the
transformation term is \cite{BOS,UJ}
\begin{eqnarray} \label{rgt}
K &=& \exp \{ -a(\bar \psi'_{x'} - \frac{b}{n^{d}}
\sum_{x \in x'} \bar \psi_{x})
(\psi '_{x'} - \frac{b}{n^{d}} \sum_{x \in x'} \psi_{x}) \} \\
& \propto & \int D \bar \eta D \eta \exp \{ \bar \eta_{x'} ( \psi'_{x'}-
\frac{b}{n^{d}} \sum_{x \in x'} \psi_{x})+(\bar \psi'_{x'} - \frac{b}{n^{d}}
\sum_{x \in x'} \bar \psi_{x} ) \eta_{x'} + \frac{1}{a} \bar \eta_{x'}
\eta_{x'} \} \nonumber
\end{eqnarray}
where $\bar \eta , \ \eta $ are auxiliary Grassmann fields on the sites
$x'$ of a coarse lattice and $x \in x'$ extends over $n^{d}$ blocks.
$b$ is a renormalization parameter.
The second representation displays that for the case of an exact
$\delta $ function RGT, $a \to \infty $,
the transformation term is chirally invariant,
whereas for finite $a$ (smeared $\delta $ function) it breaks chiral
symmetry explicitly.

U.-J. Wiese \cite{UJ,Dallas} has applied this transformation to a
general bilinear ansatz  for the action, namely:
\begin{equation} \label{bili}
S [ \bar \psi , \psi ] = \sum_{x,y} [ i \sum_{\mu} \rho_{\mu}(x-y) \bar
\psi_{x}
\gamma_{\mu} \psi_{y} + \lambda (x-y) \bar \psi_{x} \psi_{y} ]
\end{equation}
where $\rho_{\mu}, \ \lambda $ are arbitrary functions; the only assumption
here is lattice translational invariance. The recursion relations for
these functions in an RGT can be
calculated analytically and yield the same structure as in the
bosonic case, which had been analyzed by Bell and Wilson
a long time ago \cite{BW}.
If we insert Wilson fermions as initial action, it
turns out that only $b=n^{(d-1)/2}$ leads to
a non trivial fixed point, in accordance with a dimensional consideration.
In the fixed point, the quantities: $
\alpha_{\mu}(p) \doteq \rho_{\mu}(p)/ [\rho^{2}(p) + \lambda^{2}(p)] ;
\ \beta (p) \doteq \lambda (p)/ [\rho^{2}(p) + \lambda^{2}(p)] $
take the form:
\begin{equation}
\alpha_{\mu}^{*}(p) = \sum_{\ell \in Z^{d}} \frac{p_{\mu}
+ 2 \pi \ell_{\mu}}
{(p+2\pi \ell )^{2}} \prod_{\nu } \Big( \frac{ \sin (p_{\nu}/2)}
{p_{\nu}/2 + \pi \ell_{\nu}} \Big)^{2} \quad ; \quad
\beta^{*}(p) = \frac{n}{(n-1)a}
\end{equation}
Wilson's parameter $r$, i.e. the initial chiral symmetry breaking,
disappears and the doublers do not return.
Hence for $a \to \infty $ the FPA is chirally invariant.
However, numerical studies show that in this case it is non local,
therefore there is no contradiction with the Nielsen Ninomyia theorem.
For any {\em finite} $a$ the FPA becomes {\em local},
but chiral symmetry is broken.
It turns out that for $a \simeq 4$ the locality is optimal.
\footnote{This has been shown analytically for $d=1$ and numerically
for $d=2$. It coincides
with the optimal value that Bell/Wilson and Hasenfratz/Niedermayer
found in their treatment of the free scalar fields and of the
non linear $\sigma $ model, respectively \cite{BW,HN}.

Note also that the Kadanoff transformations have properties of an
Abelian semigroup only for $a \to \infty $. Therefore it is natural
that for finite $a$ the FPA depends on the blocking factor $n$.}

The disappearance of Wilson's term can be easily understood from the fact
that it represents a (discretized) second derivative, i.e. a quadratic
momentum. In the FPA only the leading order of $\rho_{\mu}(p)$ survives
and for Wilson fermions this order is linear. For dimensional reasons,
the latter is true for all reasonable fermionic actions.
\footnote{Analogously to Bell/Wilson we could generalize the power of
the leading momentum order to $1+\varepsilon $. Then we arrive
at a non trivial fixed point for $b=n^{(d-1-\varepsilon )/2}$.
For $\varepsilon \neq 0$ these fixed points are non local, like
the initial action. Bur this case is quite artificial.}
Instead of only Wilson's term we might add discretized derivatives
of any higher orders with arbitrary coefficients. They all disappear
in the FPA and for all even orders the effect is the same as for Wilson's
term (elimination of the doublers but explicit breaking of the chiral
symmetry in the initial action), whereas for all odd orders ($>1$)
chiral symmetry persists, but the doublers too.

We arrive at exactly the same FPA if we insert SLAC or Rebbi
fermions for the initial action. This holds for all lattice fermions
which coincide in the leading order of $G^{-1}(p)$ with the naive
lattice fermions and which avoid doubling by non locality.

We arrive again at the same FPA if we insert the ansatz of
Stamatescu and Wu as initial action. We may even generalize also
this approach by using  an arbitrary real parameter $q$ for the unitarity
violation:
\begin{equation}
S_{SW} = \sum_{x} \Big\{ \sum_{\mu} [q \bar \psi_{x}
\gamma_{\mu} \psi_{x+\hat \mu}
- (1-q )\bar \psi_{x+\hat \mu} \gamma_{\mu} \psi_{x}
+ (1-2q) \bar \psi_{x} \gamma_{\mu} \psi_{x} ]
+m \bar \psi_{x} \psi_{x} \Big\}
\end{equation}
For $q = \frac{1}{2} $ we obtain naive fermions
and $q = 0, \ 1$ are the one sided
derivatives discussed by Stamatescu and Wu. For
every $q \neq \frac{1}{2}$ doubling is avoided, but unitarity
violated. But any $q \neq \frac{1}{2}$ leads to the same fixed point
as Wilson fermions, in particular unitarity is recovered there \cite{SWcom}.

If we start from naive or smeared fermions, the doublers are still present
in the FPA, i.e. here we don't arrive at a useful result by iterating
the RGT (\ref{rgt}).

Not described by the ansatz (\ref{bili}) are e.g. staggered fermions,
because their action has not the full lattice
translational invariance. However, they can be treated similarly
if the RGT is modified \cite{WUJ,Dallas}. At the end, corners of hypercubic
blocks (pseudoflavors) are to reconstruct spinors, therefore
it is important not
to mix them in the RGT. Kalkreuter, Mack and Speh
have proposed a suitable transformation,
where each pseudoflavor builds its own block spin
with blocking factor 3 \cite{KMS}.
We have first applied the exact $\delta $ RGT
to this blocking scheme
and observed that the fixed point -- which is reached
for the suitable value of the renormalization parameter -- is local.
In analogy to the chiral invariance of (\ref{rgt}) for $a \to \infty $,
here the $U(1) \otimes U(1)$ symmetry for $m=0$ -- the remainder of the
chiral symmetry -- still holds in the FPA.
There is no contradiction with the Nielsen Ninomyia theorem, since
this symmetry does not imply full chiral invariance of the spinors,
which can be reconstructed. We can even smear the $\delta $ RGT
without loss of any symmetry by adding a ``kinetic term of the auxiliary
fields'' instead of the mass term in (\ref{rgt}) (the latter would
break the $U(1) \otimes U(1)$ symmetry) and obtain in this way an
extremely local FPA.

If we denote the pseudoflavors as $\bar \chi^{(i)}, \chi^{(i)}, \
i=1 \dots 2^{d}$, then the general ansatz for the staggered
fermion action -- analogous to (\ref{bili}) -- reads:
\begin{equation}
S[\bar \chi , \chi ] = i \sum_{x,y} \sum_{i,j} \bar \chi^{(i)}_{x}
\rho_{ij}(x-y) \chi^{(j)}_{y}
\end{equation}
where $x,y$ refer now to the block centers (spacing 2). After exploiting
the symmetry properties, we are left with $d$ independent elements
of $\rho $.

In particular for $d=2$ and $\bar \chi^{(i)}_{x},\chi^{(i)}_{x} =
\bar \chi_{x+a_{i}}, \chi_{x+a_{i}},\ a_{i} = (n_{1}-\frac{1}{2})\hat 1 \\
+(n_{2}-\frac{1}{2}) \hat 2$ with $i=1+n_{1}+2n_{2}$, $\rho $ can be written
as:
\begin{eqnarray}
\rho (p) &=& \frac{1}{\alpha_{1}(p)^{2}+\alpha_{2}(p)^{2}}
\left( \begin{array}{cccc} 0 & \alpha_{1}(p) & \alpha_{2}(p) & 0 \\
\alpha_{1}(p) & 0 & 0 & -\alpha_{2}(p) \\ \alpha_{2}(p) & 0 & 0 &
\alpha_{1}(p) \\
0 & -\alpha_{2}(p) & \alpha_{1}(p) & 0 \end{array} \right) \nonumber \\
&=& \frac{1}{\alpha_{1}(p)^{2}+\alpha_{2}(p)^{2}} \ \rho (p)^{-1}
\end{eqnarray}
and in the fixed point we obtain:
\begin{equation}
\alpha_{\mu}^{*}(p) = 2 \sum_{\ell \in Z^{2}} \frac{p_{\mu}+2\pi \ell_{\mu}}
{(p+2\pi \ell )^{2}} (-1)^{\ell_{\mu}} \prod_{\nu}
\Big( \frac{\sin (p_{\nu}/2)}{p_{\nu}/2+\pi \ell_{\nu}} \Big)^{2} +
\frac{9}{8a} \sin (p_{\mu}/2)
\end{equation}
where the auxiliary kinetic term is suppressed by a factor $1/a$.
Optimal locality is reached for $a \simeq 9/4$, as we see again
analytically for $d=1$ and numerically for $d=2$.

We also treated Wilson fermions with a weak four fermion interaction
in the framework of the Gross Neveu model \cite{Dallas}.
There we studied for $d=2$ the vicinity of the non local fixed
point mentioned above. The interaction was expressed by an auxiliary
scalar field $\phi $ with a Yukawa coupling $y$. The general ansatz
for the Yukawa coupling after a number of RGT reads:
\begin{displaymath}
\frac{1}{(2\pi )^{4}} \int dp dk \ \bar \psi (p) \sigma (p,k) \psi (k)
\phi (-p-k)
\end{displaymath}
The RGT also includes a blocking of $\phi $ to coarser lattices.
At the fixed point, the matrix $\sigma  $ takes to the first order
of $y$ the form:
\begin{eqnarray}
\sigma^{*}(p,k) &=& y \rho^{*}_{\mu}(p) \gamma_{\mu}
\sum_{\ell , l \in Z^{2}}
\frac{p_{\nu}+2\pi \ell_{\nu}}{(p+2\pi \ell )^{2}} \gamma_{\nu}
\frac{k_{\rho}+2\pi l_{\rho}}{(k+2\pi l )^{2}} \gamma_{\rho}
\rho^{*}_{\sigma}(k) \gamma_{\sigma} \nonumber \\
&& \prod_{\lambda} \frac{2 \sin ((p_{\lambda}+k_{\lambda})/2)}
{p_{\lambda}+2\pi \ell_{\lambda} + k_{\lambda} + 2\pi l_{\lambda}}
\frac{2 \sin (p_{\lambda}/2)}{p_{\lambda} + 2\pi \ell_{\lambda}}
\frac{2 \sin (k_{\lambda}/2)}{k_{\lambda} + 2\pi l_{\lambda}}
\end{eqnarray}
For interacting fermions there is a generalization of the No Go
theorem due to Pelissetto \cite{Pelissetto}, which applies
to certain non local theories involving a singularity of $G^{-1}(p)$.
\footnote{The type of non local lattice fermions with a finite gap
of $G^{-1}(p)$ -- such as SLAC fermions -- is also troublesome on the loop
levels; for this case there are problems to recover
Lorentz invariance in the continuum \cite{KaSi}.}
For Rebbi fermions, which have
some similarity with our non local fixed point, it turned out
that there are still doublers beyond the tree level \cite{mafia}.
We refer to spurious ghost states which are dynamically generated.
It remains to be checked if our FPA suffers from this problem too.
There is hope that this is not the case since in our case the
poles in $G^{-1}(p)$ appear naturally from the RGT, unlike those
which Rebbi has created by hand.

The analogous consideration
for staggered fermions is in progress. Since the Gross Neveu model is
asymptotically free, there is one (weakly) relevant direction
(to lowest order it is marginal),
which can be determined perturbatively. This is the tangent to
the curve of perfect actions -- actions free of cutoff effects --
that emanates from the FPA, away from the critical surface.
The concept of Hasenfratz
and Niedermayer \cite{HN} consists now of following this tangent to a
point where the correlation length is short enough for simulations
and hoping to be still close to a perfect action.
Then continuum physics can be described to a very good approximation
with only little numerical effort.

The final goal of this concept is to improve the scaling of QCD.
As a preparatory work for the combination with gauge theory,
we considered the pure Schwinger model with a simple $\delta $
RGT -- that just sums over the frames of hypercubic blocks --
and found also there analytically a
similar recursion relation and a fixed point \cite{WUJ}. To smear the
$\delta $ RGT, however, a
more sophisticated blocking scheme is required. Probably it will be necessary
to include all the gauge fields on the fine lattice in the RGT.
We note that if we switch on a gauge interaction between fermions, the
``kinetic smearing'' of the $\delta $ RGT is not permitted any more.\\

{\em Acknowledgement} \ \ It is a pleasure to thank U.-J. Wiese
for our stimulating collaboration and for introducing me to
to the issue of lattice fermions.

Moreover we  are indebted to P. Hasenfratz, F. Niedermayer,
A. Pelissetto, and I. Stamatescu for interesting discussions.


\begin{thebibliography}{20}
\bibitem{SWcom}
Stamatescu and Wu motivate their new animal in the
lattice fermion zoo with a postulate that forbids any continuous
parameters which disappear in the continuum limit -- except for
the lattice spacing -- such as Wilson's parameter $r$, because
``irrelevant'' parameters can play quite a relevant role sometimes.
{}From our point of view,
their fermion type deserves attention, since it leads to a unitary
fixed point, but we would recommend to drop the justification from
this postulate. Firstly the postulate itself is rather dubious
and secondly, as we illustrate above, their own fermions do not really
respect it. The authors have just hidden the possibility of such a parameter
$q$ by focusing only on particular values of it. Similarly we might
set Wilson's $r=1$ and claim that there is no continuous parameter
involved. In their last appendix, Stamatescu and Wu mention
the possibility of a continuous unitarity breaking parameter,
but reject it arguing with a consideration on stationary points
of the dispersion law. We are not convinced by this argument.
\end{thebibliography}
\end{document}